\documentclass[aps,amssymb,twocolumn,superscriptaddress,nofootinbib]{revtex4}
\usepackage{setspace}
\usepackage{graphicx}
\usepackage{psfrag}

\newcommand{\Screll}{{\mathcal L}}

\def\be{\begin{equation}}
\def\ee{\end{equation}  }
\def\bea{\begin{eqnarray}}
\def\eea{\end{eqnarray}  }
\def\rg{\sqrt{-g}}

\begin{document}
\title{Simulation of Binary Black Hole Spacetimes with a Harmonic Evolution Scheme}

\author{Frans Pretorius}
\affiliation{Department of Physics,
             University of Alberta,
             Edmonton, AB T6G 2J1 }
\affiliation{Canadian Institute for Advanced Research, Cosmology and Gravity Program}
\begin{abstract}
A numerical solution scheme for the Einstein field equations
based on generalized harmonic coordinates is described, focusing on details not
provided before in the literature and that are of particular relevance
to the binary black hole problem. This includes demonstrations of the
effectiveness of constraint damping, and how the time slicing can be 
controlled through the use of a source function evolution equation.
In addition, some results from an ongoing study of binary black hole coalescence,
where the black holes are formed via scalar field collapse, are shown.
Scalar fields offer a convenient route to 
exploring certain aspects of black hole interactions, and one interesting,
though tentative
suggestion from this early study is that behavior reminiscent of ``zoom-whirl'' 
orbits in particle trajectories is also present in the merger of equal
mass, non-spinning binaries, with appropriately fine-tuned initial conditions.
\end{abstract}

\maketitle

\section{Introduction}
The past year has seen a remarkable leap in progress toward a numerical
solution of the binary black hole problem. The first stable evolution
of an entire merger process---orbit, merger, ringdown and gravitational
wave extraction---was presented in~\cite{paper2} using a harmonic
formulation of the field equations.
Shortly afterwards two groups~\cite{utb,nasa} independently achieved similar
success using a modified form of the BSSN (or NOK)~\cite{nok,bs,sn} formulation of the
field equations. These first results all focused on the merger of
equal mass, non-rotating black holes, and follow-up studies~\cite{utb2,nasa2}
are now honing in on a consistent picture of the gravitational waves
emitted by such an event. Recently, similar techniques to those employed 
in~\cite{utb,nasa} were successfully used to 
study unequal mass mergers~\cite{penn,nasa3} and provide estimates
of the ``kick'' velocity imparted to the final black hole, and 
in~\cite{utb3} the effects of black hole spin, aligned and anti-aligned
with the orbital angular momentum, were studied, demonstrating that
in the aligned case some orbital ``hang-up'' occurs to radiate away
the excess angular momentum.

One reason why the new
results seem like such a leap is that progress 
during earlier years had been rather slow and arduous.
For example, it had taken roughly 7 years from the first simulation
of a fraction of a non head-on collision~\cite{bruegmann} to a full orbit\cite{bruegmann_et_al},
with many groups making advances along the way. Much of the
effort was also (and still is) focused on understanding the underlying
structure of the field equations of general relativity, and why
they are so problematic to discretize successfully in many cases. Nevertheless,
the manner in which advances had been made over the pass decade suggested
a similar series of ``baby steps'' toward a future solution, which
is why the results of the past year have been so exciting.

The primary purpose of this paper is to describe in detail certain 
aspects of the generalized harmonic (GH) evolution scheme that are of 
relevance to the simulation of binary black hole (BBH) spacetimes, 
as first presented in~\cite{paper1,paper2}. This is not
a comprehensive overview of the code or technique, and could be viewed as 
a supplement to~\cite{paper1,paper2} 
(and see also~\cite{new_lindblom_et_al} for an excellent description
of GH evolution). A secondary goal
is to relate progress on an ongoing effort to understand the 
nature of BBH
spacetimes produced by scalar-field collapse\footnote{a study of 
Cook-Pfeiffer quasi-circular inspiral initial data\cite{cook_pfeiffer} 
evolved using the generalized
harmonic evolution scheme will be presented elsewhere\cite{buonanno_et_al}}. 
It is arguable how relevant such spacetimes may eventually be to 
gravitational wave detection efforts,
nevertheless they offer an easy route to explore a wide range 
of BBH parameter space. One interesting though tentative result
from this early study is that the zoom-whirl type behavior seen
in geodesics around black holes may also be present in BBH
orbits involving comparable mass components. In particular,
what is shown is that for equal mass binaries on non-circular 
orbits, and with appropriate fine-tuning of the initial
conditions, the black holes approach one another, ``whirl'' around
for several orbits, then separate again.

The outline of the rest of the paper is as follows. In Sec.~\ref{sec_method} 
a brief overview of the Einstein field equations in harmonic form with 
dynamically evolved {\em source functions} and {\em constraint damping} terms
is given.
Sec.~\ref{sec_num_details} contains a description of parts of the numerical code that
seem to be important for stable evolution of BBH spacetimes with this scheme, including
excision and numerical dissipation. The effects of constraint damping and the choice
of source function evolution equations on a select set of simulations is presented
in Sec.~\ref{sec_cdg}. Results from the inspiral and merger of scalar field generated
BBH spacetimes are given in Sec.~\ref{sec_bbh}, followed by concluding remarks 
in Sec.~\ref{sec_conclusion}.

\section{Generalized Harmonic Coordinates}\label{sec_method}

In this section a brief description of generalized harmonic coordinates,
in the specific form that is discretized in the code, is given. 
For a thorough description
and the history of the use of these coordinates see~\cite{new_lindblom_et_al}.

Consider a spacetime describe by the line element $ds$, metric tensor $g_{ab}$
and coordinates $x^a$
\be
ds^2=g_{ab} dx^a dx^b,
\ee
and the Einstein field equations in the form
\be\label{efe}
R_{ab}=8\pi\left(T_{ab}-\frac{1}{2}g_{ab} T\right),
\ee
where $R_{ab}$ is the Ricci tensor, $g_{ab}$ is the
metric tensor, $T_{ab}$ is the stress energy tensor
with trace $T$, and units have been chosen so that Newton's constant $G$
and the speed of light $c$ are equal to 1. 
The Ricci tensor is defined in terms of the Christoffel
symbols $\Gamma^{c}_{ab}$ 
\be
\Gamma^{c}_{ab}=\frac{1}{2}
g^{ce}\left[g_{ae,b}+g_{be,a}-g_{ab,e}
\right]
\ee
via
\be\label{ricci}
R_{ab}=\Gamma^{d}_{ab,d} - 
       \Gamma^{d}_{db,a}
      +\Gamma^{e}_{ab}\Gamma^{d}_{ed}
      -\Gamma^{e}_{db}\Gamma^{d}_{ea}
\ee
The notation $f_{,a}$ and $\partial_{a}f$ is used interchangeably to
denote ordinary differentiation of some quantity $f$ with respect to the coordinate
$x^a$. 

{\em Harmonic coordinates} are defined by the following set of four conditions
on the four spacetime coordinates:
\be\label{harm_def}
\Box x^c = 0,
\ee
where $\Box$ is the usual covariant scalar wave operator:
\be
\Box x^c = \nabla_b\nabla^b x^c = \frac{1}{\rg}\partial_a\left(\rg g^{ac} \right).
\ee
{\em Generalized harmonic (GH) coordinates} introduce a set of {\em arbitrary
 source functions} $H^c$ into the definition (\ref{harm_def}):
\be\label{gharm_def}
\Box x^c = H^c.
\ee
Note that in contrast to harmonic coordinates (\ref{harm_def}), 
GH coordinates (\ref{gharm_def}) are {\em not} conditions
on the spacetime coordinates---any geometry in any coordinate
system can be expressed in GH form with (\ref{gharm_def}) defining
the corresponding source functions.

Using the definitions above, and defining
$H_c=g_{ac}H^a$, the Einstein
equations (\ref{efe}) can be rewritten in the following equivalent
form, which shall be referred to as the generalized harmonic decomposition
of the field equations:
\bea
\frac{1}{2} g^{cd}g_{ab,cd} 
+ g^{cd}{}_{(,a} g_{b)d,c}
+ H_{(a,b)} \nonumber
- H_d \Gamma^d_{ab} \\
+ \Gamma^c_{bd}\Gamma^d_{ac} 
= - 8\pi\left(T_{\alpha\beta}-\frac{1}{2}g_{\alpha\beta} \label{efe_h}T\right)
\eea
The utility of the GH decomposition in a numerical 
evolution (as first carried out in~\cite{garfinkle,szilagyi_et_al,szilagyi_winicour},
and also~\cite{babiuc_et_al1,new_lindblom_et_al,babiuc_et_al}) comes
from considering the $H_c$ as {\em independent functions};
(\ref{efe_h}) then becomes a set of ten manifestly hyperbolic equations
for the ten metric elements $g_{ab}$.
As $H_c$ are now four independent functions,
one needs to provide four additional, independent differential equations to 
solve for them, schematically written as
\be\label{he}
\Screll_c H_c = 0 \ \ \ \mbox{(no summation)}.
\ee
$\Screll_c$ is a differential operator that in general depends
on the spacetime coordinates, metric and source functions. To complete
the specification of the system one needs to provide evolution equations
for matter sources, which couple to the field equations through
the stress energy tensor. The only matter field considered here
is a massless scalar field $\Phi$ that satisfies the wave equation
\begin{equation}\label{phi_eom}
\Box \Phi = 0,
\end{equation}
and has a stress energy tensor
\begin{equation}\label{set}
T_{ab} = 2 \Phi_{,a}\Phi_{,b} - g_{ab} \Phi_{,c}\Phi^{,c},
\end{equation}
A solution to (\ref{efe_h}), (\ref{he}) and (\ref{phi_eom})
will be a solution to the Einstein equations 
as long as the GH condition (\ref{gharm_def}) is satisfied for all 
time. In theory, this is straight-forward to achieve. Define the
GH constraint functions $C^a$ by
\be
C^a \equiv H^a - \Box x^a. \label{c_def}
\ee
Any solution to the Einstein equations must have $C^a=0$. 
Using the contracted Bianchi identity and 
conservation of stress energy, one can show that $C^a$ satisfies
the following homogeneous wave equation 
\be\label{h_const}
\Box C^a = - R^a{}_b C^b.
\ee
Therefore, what needs to be done to ensure that (\ref{efe_h}), (\ref{he}) 
and (\ref{phi_eom})
satisfy the Einstein equations for all time $t$ is to choose initial
conditions for $g_{ab}$ and $H_a$ such that $C^a=0$ and 
$\partial_t C^a=0$ at $t=0$, boundary conditions on $g_{ab}$ and $H_a$
such that $C^a=0$ on the boundary for all time, and couple this to
matter that conserves stress energy. Then (\ref{h_const}) guarantees
that $C^a=0$ throughout the interior 4-volume of the spacetime.

Of course, things are not as simple as this in practice. In a numerical
simulation one can only satisfy the conditions described
in the preceding paragraph to within the truncation error of the
numerical scheme. Again, in principle this is not a problem, however 
it turns out that in many situations 
the truncation errors grow too rapidly to achieve
useful results given limited resolution.
There is good evidence that
one of the reasons for the rapid growth of truncation error is
{\em not} a poor choice of a numerical algorithm, rather (\ref{h_const})
admits rapidly growing solutions (so called ``constraint violating modes'')
given initial data where $C^a$ is of order the truncation error. An
effective way of dealing with this problem is the addition of 
{\em constraint damping} terms to the equations.

\subsection{Constraint Damping}

As introduced by Gundlach et al.\cite{gundlach_et_al}, building
on the idea of so-called $\lambda$ systems~\cite{lambda_ref},
the field equations with constraint damping are
\bea
\frac{1}{2} g^{cd}g_{ab,cd} 
+ g^{cd}{}_{(,a} g_{b)d,c}
+ H_{(a,b)} \nonumber
- H_d \Gamma^d_{ab} \\
+ \Gamma^c_{bd}\Gamma^d_{ac} 
+ \kappa\left(n_{(a} C_{b)} - \frac{1}{2}g_{ab} n^d C_d\right) \nonumber\\
= - 8\pi\left(T_{\alpha\beta}-\frac{1}{2}g_{\alpha\beta} T\right),\label{efe_h_cd}
\eea
where $\kappa$ is a parameter multiplying the new terms,
$n_a$ a timelike vector, and $C_a$ are the constraints (\ref{h_const}). Since
the new terms are proportional to the constraints, a solution to the Einstein
equations (\ref{efe}) will also be a solution to (\ref{efe_h_cd}); furthermore,
for general solutions of (\ref{efe_h_cd}) the constraints still satisfy
a homogeneous wave equation
\be
\Box C^a = - R^a{}_b C^b + 2\kappa\nabla_b\left[n^{(b}C^{a)}\right],\label{new_cp}
\ee
and thus the prescription outlined in the previous section for obtaining valid 
solutions to the Einstein equations can also be applied using (\ref{efe_h_cd})
instead of (\ref{efe_h}). Gundlach et al. have shown
that for perturbations about Minkowski spacetime, all finite wavelength 
solutions to (\ref{new_cp}) are exponentially damped in time.
It is not known whether similar 
modifications to other forms of the field equations can be made
that will also have this desirable constraint damping property, though in~\cite{gundlach_et_al}
it was shown how to apply constraint damping to the $Z4$ formalism~\cite{z4}, and recently 
Lindblom et al.\cite{new_lindblom_et_al} described a first order symmetric hyperbolic
version of the GH decomposition with constraint damping. In a first order
form of the equations additional constraints are introduced 
that could also exhibit poor behavior with regards to being satisfied in a numerical
evolution, though~\cite{new_lindblom_et_al} describe an effective method
for dealing with this problem.

In~\cite{gundlach_et_al} it was suggested that $n^a$ in (\ref{efe_h_cd})
could be any timelike vector field; here, for simplicity, $n^a$ is chosen
to be the unit timelike vector normal to $t\equiv x^0 = {\rm const.}$ 
surfaces. Specifically, $n_a = -\alpha \partial_a t$, with $\alpha=\sqrt{-1/g^{tt}}$ 
being the usual {\em lapse function}.

\subsection{Source Function Evolution}
Within the GH decomposition one can think
of the source functions $H_a$ as representing the four
coordinate degrees of freedom available in general relativity.
There are many conceivable ways of choosing $H_a$---see~\cite{paper1}
for a more general discussion of some possibilities, and~\cite{friedrich2,balakrishna_et_al,alcubierre_et_al,lindblom_scheel,alcubierre_et_al_2,bona_et_al,bona_et_al_2} 
for coordinate conditions that might be readily
applicable to GH evolution.
The focus here will be on one class of equations that have
proven useful for binary black hole evolutions, namely:
\be
\Box H_t = - \xi_1 \frac{\alpha-1}{\alpha^\eta} + \xi_2 H_{t,\nu} n^\nu\label{t_gauge}, \ \ \ H_i=0.
\ee
This equation 
for $H_t$ is a damped wave equation with forcing function
designed to prevent the lapse from deviating too far from its Minkowski value of 
$1$. The parameter $\xi_2$ controls the damping term,
and $\xi_2,\eta$ regulate the forcing term. The GH numerical code descrided
here tends to develop instabilities in pure harmonic gauge when the lapse function becomes
on the order of $0.1$ near the horizon of black 
holes; equation (\ref{t_gauge}) prevents this from happening. It is unclear
whether the instabilities are numerical in nature or an indication that
the harmonic gauge is developing a coordinate pathology.
However, this
is not a crucial question to answer at the moment given the limited 
availability of computational resources to 
fully explore the issue, and that (\ref{t_gauge}) works well.
In Sec.~\ref{sec_cdg} examples of the effect 
of (\ref{t_gauge}) with a few different parameters are shown.

\subsection{Boosted Scalar Field Initial Data}

An easy way to produce binary black hole ``initial'' data is to use 
scalar field gravitational collapse. 
At $t=0$ one begins with two Lorentz boosted scalar field profiles with
initial amplitude, separation and boost parameters
to approximate the kind of orbit that the black
holes, which form as the scalar field collapses, will have.
On roughly the light crossing time scale of the orbit the
remnant scalar field that did not fall into either black hole propagates
away from the vicinity of the orbit, leaving behind a good approximation
to a vacuum BBH system.

The procedure used to calculate the initial geometry is
based on standard techniques~\cite{cook}. One starts
with a metric written in ADM~\cite{ADM} form
\be\label{adm}
ds^2 = -\alpha^2 dt^2 + h_{ij} \left(dx^i + \beta^idt\right)\left(dx^j + \beta^jdt\right),
\ee
where as before $\alpha$ is the lapse, $\beta^i$ the shift vector
and $h_{ij}$ is the metric intrinsic to $t={\rm const.}$ surfaces. 
The extrinsic curvature of $h_{ij}$ is defined as
\be
K_{ij}= - h_i{}^{l} h_j{}^{m} \nabla_m n_l.\label{kdef}
\ee
In the ADM decomposition
the Hamiltonian and momentum constraint equations take on the
following form:
\bea
^{(3)}R + K^2 - K_{ab} K^{ab} = 16\pi\rho, \label{hc} \\
\nabla_b K_a{ }^{b} -\nabla_a K = 8\pi J_a \label{mc},
\eea
where $^{(3)}R$ is the trace of the Ricci tensor of the
spatial metric, $K$ is the trace of the extrinsic curvature,
$\rho$ is the energy density 
and $J_a$ the momentum density of the matter in the spacetime:
\bea
\rho &=& T_{ab} n^a n^b, \\
J_a &=& - T_{lm} h^l{}_a n^m. \label{stress}
\eea
For the results described here
the following initial conditions for the geometry
are chosen at $t=0$: the spatial
metric and its first time derivative is conformally flat:
\bea
h_{ij}|_{t=0} = \psi \cdot \eta_{ij}\label{cflat}\\
\partial_t h_{ij}|_{t=0} = \partial_t \psi \cdot \eta_{ij}
\eea
where $\psi$ is the conformal factor, and $\eta_{ij}$ is 
the flat Euclidean metric; 
the initial slice is {\em maximal}:
\be
K=0\label{max_slice_a}, \ \ \
\partial_t K=0 \label{max_slice}
\ee
and the initial slice is harmonic:
\be
H_t=0, \ \ \ H_i=0\label{ic_harm}.
\ee

When the conditions (\ref{cflat}-\ref{max_slice}) are substituted into
the constraints (\ref{hc}-\ref{mc}), the Hamiltonian constraint
becomes an elliptic equation for the conformal factor $\psi$, and
the momentum constraints become elliptic equations for the
components of the shift vector $\beta^i$. When the
maximal slicing condition (\ref{max_slice}) is expanded in terms
of the metric via its definition (\ref{kdef}), and using
the above conditions, an elliptic equation for the lapse $\alpha$
results. This set of five coupled elliptic equations is solved
using multigrid techniques as described in~\cite{paper1}. For boundary
conditions the metric is assumed to be the Minkowski metric (and these
conditions are applied {\em exactly}, as a coordinate system
compactified to spatial infinity is used).
For scalar field collapse initial data no ``inner''
boundary conditions are needed as there are no black holes present
in the initial slice. After the elliptic equations are solved
for $\alpha,\psi$ and $\beta^i$, (\ref{max_slice_a}) is algebraically
solved for the initial value of $\partial_t \psi$, and similarly
the conditions (\ref{ic_harm}) together with the definition
(\ref{gharm_def}) is used to solve for
the initial values of $\partial_t \alpha$ and $\partial_t \beta^i$. 
Once all the initial values and first time derivatives of the
metric in ADM form (\ref{adm})
are known, it is straight-forward to 
convert these to initial conditions for the 4-dimensional metric
$g_{ab}$ needed for the GH system of equations (\ref{efe_h_cd}).
In addition to (\ref{ic_harm}), the first time derivative of
$H_t$ is needed for initial conditions to (\ref{t_gauge}), and this is
chosen to be:
\be
\partial_t H_t=0.
\ee
Note that the above procedure produces initial data that is consistent
with the GH constraints (\ref{c_def}) $C^a=0$, and $\partial_t C^a=0$
at $t=0$: $C^a=0$ explicitly as this definition is used to provide
the initial time derivatives of the lapse and shift, and 
initial data that satisfies the ADM constraints (\ref{hc}-\ref{mc})
will, to within truncation error, satisfy $\partial_t C^a=0$ 
then~\cite{friedrich}.

The scalar field initial data is constructed by first taking a spherical, time symmetric
Gaussian profile in a Minkowski rest frame $ds^2=-dt'^2+dx'^2+dy'^2+dz'^2$:
\bea
\Phi(t'=0,x',y',z') &=& A \exp\left(-\frac{r'^2}{\Delta^2}\right) \nonumber\\
\partial_{t'} \Phi(t'=0,x',y',z') &=& 0,\label{sf_id}
\eea
where $A$ and $\Delta$ are constant parameters, and $r'=\sqrt{x'^2+y'^2+z'^2}$.
Then, a Lorentz boost with velocity $v$ is applied in the direction
$\vec{u}$ to this pulse, mapping $(t',x',y',z')=(0,0,0,0)$
to $(t,x,y,z)=(0,x_0,y_0,z_0)$, where $(x_0,y_0,z_0)$ is the initial
location of the pulse in simulation coordinates.
Note that this initial data
scheme can only produce a black hole (when the amplitude $A$ is sufficiently
large) with approximately the velocity $v$, as no back reaction effects
are taken into account\footnote{for a boost parameter of order $0.2$, as
used in the simulations described in Sec.~\ref{sec_bbh},
the estimated velocities of the resultant black holes can differ
from $v$ by as much as $20-30\%$.}.

\section{Numerical Simulation with Generalized Harmonic Coordinates}\label{sec_num_details}

Here a very brief overview of the numerical code implementing the
GH system of equations (\ref{phi_eom}), (\ref{efe_h_cd}) and (\ref{t_gauge})
is given, focusing on a few technical details
and miscellaneous information related to binary black hole simulation
that are not discussed in~\cite{paper1}.
The numerical code has the following features:

$\bullet$ Equations (\ref{phi_eom}), (\ref{efe_h_cd}) and (\ref{t_gauge})
      are discretized using standard second order accurate finite difference
      stencils. The evolved variables are the ten covariant metric elements
      $g_{ab}$, the four source functions $H_a$, and the scalar field $\Phi$. 
      A three time level scheme is used, where unknown quantities
      at the most advanced time level are solved for using a pointwise 
      Newton-Gauss-Seidel relaxation, given the known quantities at the two 
      past time levels.

$\bullet$ The constraint equations (\ref{hc}) and (\ref{mc}) are solved 
      with a Full Approximation Storage (FAS) adaptive multigrid algorithm.

$\bullet$ Excision is used to eliminate the singularities inside black holes.
      The excision surface is an ellipse whose shape is a shrunken version
      of a best-fit match to the shape of the apparent horizon (AH) of the
      black hole. The ellipse is shrunk, by typically $0.5$ to $0.9$, so 
      that the excision surface is always some distance inside the AH.
      The excision surface
      is redefined each time the AH is search for, which is 
      frequently enough that the excision surface never moves by more than a single
      grid point between searches. This is to ensure stability as extrapolation is used
      to initialize newly ``repopulated'' grid points, and extrapolation tends
      to be unstable if more than a single point inward from the excision
      surface is repopulated at any one time.

$\bullet$ A spatially compactified coordinate system is used so that exact Minkowski
      spacetime boundary conditions can be placed on all variables. 

$\bullet$ A combination of adaptive and fixed mesh refinement is used to efficiently resolve
      the relevant length scales in a given simulation. The grid hierarchy is adaptive
      in the ``near-zone'' to track the
      motion of the black holes through the domain, while outside of this 
      in the ``wave-zone'' the mesh structure
      is kept fixed. This approach, rather than full adaptivity, is for computational 
      efficiency---the simulations would take a prohibitively
      long time if the mesh structure were allowed to track the outgoing wave train.
      The software libraries used to implement the parallel adaptive infrastructure
      are publicly available~\cite{pamr_amrd}, though at present the documentation 
      is sparse.

$\bullet$ Numerical dissipation is used to control high frequency ``noise'' that often
      arises in such adaptive simulations. Furthermore, dissipation is necessary
      to eliminate a high frequency instability that would otherwise occur
      near black holes, where the local lightcones start to tip in towards 
      them~\cite{calabrese_pc,babiuc_et_al1}.

The remainder of this section contains a discussion of how certain properties
of the solution are measured, and a few technical details of
the code: position dependent dissipation and constraint damping
parameters, and how some robustness problems
in the apparent horizon finder are dealt with.

\subsection{Measurement of Solution Properties}\label{sec_solp}
Here a brief summary is given of how black hole properties
are measured and gravitational waves are extracted from a numerical solution.

At present only the apparent horizons and not event horizons of
black holes are searched for in the solution. 
Black hole masses are estimated from the AH area $A$
and angular momentum $J$, and applying the Smarr formula:
\begin{equation}\label{smarr}
M=\sqrt{M_{ir}^2 + J^2/(4 M_{ir}^2)}, \ \ \ M_{ir}\equiv \sqrt{A/16\pi}.
\end{equation}
The angular momentum of horizons
are calculated using two methods.
First, by using the dynamical
horizon framework~\cite{ashtekar_krishnan}, though {\em assuming}
that the rotation axis of the black hole
is orthogonal to the $z=0$ orbital plane, and that each
closed orbit of the azimuthal vector field 
lies in a $z={\rm constant}$
surface of the simulation. Due to the symmetry of the initial data,
these assumptions are probably valid, though this will eventually need
to be confirmed. The second method, following~\cite{brandt_seidel},
is to measure the ratio $C_r$ of the polar to equatorial
proper radius of the horizon, and use the
formula that closely approximates the function for
Kerr black holes:
\begin{equation}\label{a_cr}
a \approx \sqrt{1-\left(2.55 C_r -1.55\right)^2}
\end{equation}

To calculate the gravitational waves emitted by the binary
the Newman-Penrose scalar $\Psi_4$ is used, with the null
tetrad constructed from the unit timelike normal $n^\mu$,
a radial unit spacelike vector normal to $r={\rm constant}$
coordinate spheres, and two additional unit spacelike
vectors orthogonal to the radial 
vector\footnote{At this stage 
all the subtleties in choosing an ``appropriate'' tetrad are ignored---see for 
example~\cite{kinnersley_stuff}}.
Far from the source, the
real and imaginary components of $\Psi_4$ are
proportional to the second time derivatives of the
two polarizations of the emitted gravitational
waves.

The following formula~\cite{smarr_rad} is used 
to estimate the total energy $E$ emitted in gravitational waves,
\begin{equation}
\frac{dE}{dt}=\frac{R^2}{4\pi} \int p d\Omega, \ \ \
p=\int_0^t \Psi_4 dt \cdot \int_0^t \bar{\Psi}_4 dt \label{Ep},
\end{equation}
where $\bar{\Psi}_4$ is the complex conjugate of $\Psi_4$,
and the surface integrated over in (\ref{Ep}) is
a sphere of constant coordinate radius $R$.
However, before applying this formula $\Psi_4$ is filtered
by eliminating all but the $\ell=2,m=\pm 2$ spin weight $-2$
spherical harmonic components $_{-2} Y_{\ell,m}(\theta,\phi)$,
which are the dominant modes of the gravitational wave\footnote{
note that in~\cite{paper2} the waveform shown was unfiltered,
and for the energy calculation the filtering was done
with scalar spherical harmonics}. 
This eliminates some high-frequency ``noise'' that is present
in the bare waveform (primarily from mesh-refinement effects), however the integrated energies in the 
filtered vs. unfiltered waveform do not differ by more than $5\%$
at most in a typical simulation. The plots of waveforms
given in Sec.~\ref{sec_bbh} show the dominant harmonic components
of $\Psi_4$ as a function of time, calculated over a given coordinate
sphere of radius $r$:
\be\label{psi4_c_def}
_{-2}C_{\ell,m}(t,r) \equiv \int \Psi_4(r,t,\theta,\phi)\cdot {}_{-2} \bar{Y}_{\ell,m}(\theta,\phi) \ d\Omega
\ee

\subsection{Position Dependent Dissipation and Constraint Damping}
With this evolution scheme numerical dissipation is essential
to control what would otherwise be a high-frequency instability 
near black holes---see~\cite{calabrese_pc,babiuc_et_al1}.
Also, dissipation is useful in suppressing
spurious high frequency components of the numerical solution that
are sometimes generated at mesh refinement boundaries. The
Kreiss-Oliger style dissipation employed (see~\cite{paper1}) 
converges away in the continuum limit, though with typical resolutions
used in these simulations the dissipation can cause noticeable degradation
in gravitational waves measured far from the source.
Experiments suggest a dissipation parameter $\epsilon$ ($0\leq\epsilon<1$) 
of at least $0.3-0.5$
is needed for stability near black holes, though far from the black holes
much less is needed to control mesh refinement ``noise''. Therefore,
in these simulations a position dependent dissipation parameter is used
which is as large as needed near black holes for stability, and then
drops to a smaller value in the wave zone for improved accuracy
of the gravitational waveform\footnote{the simulations shown in 
\cite{paper2} used a constant value of $\epsilon$ throughout
the domain}. Specifically, inside an AH
a value of $\epsilon=0.5$ is used on
the excision surface (which is between $50\%$ and $90\%$ the size of
the AH), then $\epsilon$ is decreased
to $0.35$ linearly with coordinate distance from the center of the AH
to its surface. Outside any AH $\epsilon$ is set to $0.35$
at $r=0$ ($r=\sqrt{x^2+y^2+z^2}$ is coordinate distance from the origin),
then linearly interpolated to $0.05$ at $r\approx10M_0$ ($M_0$ is
the sum of initial black hole masses); for $r$ greater than this 
$\epsilon$ is kept fixed at $0.05$.

In certain situations it may be necessary to use
a position dependent constraint damping parameter, i.e. redefine
$\kappa$ in (\ref{efe_h_cd}) to be $\kappa(x,y,z)$. As demonstrated in
Sec.~\ref{sec_cdg}, $\kappa$ needs to be of order $1/L$, where
$L$ is the smallest lengthscale present, to be effective. However,
if $\kappa$ is much greater than 1 numerical instabilities may develop near
the outer boundaries of the domain. In those situations the following
function for $\kappa(x,y,z)$ can cure the problem:
\be
\kappa(x,y,z)=\kappa_0 \left[(1+x^2)(1+y^2)(1+z^2)\right]^{-m/2},
\ee
where $m$ is a positive integer and $\kappa_0$ is a positive constant.
For all results presented here $m=0$.

\subsection{Excision and Apparent Horizons}\label{sec_eah}

Black hole excision is used to deal with the physical singularities
that occur inside of black holes. Excision is the placement of
an artificial boundary around the singularities, though inside each black hole.
It is possible to do so because causality will prevent these unphysical interior
boundaries from affecting the exterior solution. In the code described here
all characteristics of the differential equations
being solved are {\em assumed} 
to be directed into the boundary. Thus, no boundary
conditions are placed on the excision surface; rather, the difference
equations are solved there using modified finite difference operators
that do not sample points in the computational domain that
are excised (see~\cite{paper1} for details). 

The surface that defines the excision boundary is
guided by the apparent horizon as described at the beginning of
Sec.~\ref{sec_num_details}.
An AH is search for with sufficient frequency that
the black hole does not move by more than a single grid point
in any direction between AH searches. A flow method is used to
find the AH, using the shape
of the previously found AH as an initial condition. In the
flow method a tolerance $\tau_0$ is chosen, and the
flow equation is iterated until $|\theta|_{\ell_2}\leq \tau_0$,
where $|\theta|_{\ell_2}$ is the $\ell_2$
norm of the expansion of the AH. 
This works well 
during most of the evolution, though near the time of the merger 
the flow method tends to become unstable.
If an AH is ``lost'' the simulation crashes shortly afterwards as 
the excision surface is unable to track the motion of the black hole,
and soon a domain exterior to the black hole is excised.
One of the factors that seems to cause this behavior in the AH
finder is relatively poor resolution
of the underlying numerical solution. So one cure would be to allow
additional refinement about the black holes. However, experiments
have shown that increasing the resolution only near the
black holes does not increase the accuracy in the overall 
solution, so this would be a computationally expensive solution merely 
to help the
AH finder. Of course, a better solution would be to develop a more
robust AH finder (see~\cite{thornburg,thornburg2} for a review of methods), and
this path will be pursued in future work.

In leu of a more robust AH finder the
following ``tricks'' are used to push the evolution through the merger point.
When the flow method becomes unstable, what typically happens is
$|\theta|_{\ell_2}$ decreases to some minimum value $\tau_m>\tau_0$,
then begins to increase and eventually diverge. If $\tau_m$ is not too 
much larger than $\tau_0$\footnote{typically
if $\tau_m<50\tau_0$, where the factor of $50$ was chosen
from experiments in
``normal'' situations showing that the corresponding AH shape differs from the
actual AH shape by at most around $20\%$ in size},
the surface corresponding to $|\theta|_{\ell_2}=\tau_m$ is used and evolution
continues. 
On occasion just before a merger $\tau_m$ {\em does}
become too large; in that case the motion of the AHs are extrapolated 
until an encompassing AH is found, using 
previously measured angular and radial velocities of the AHs.
This extrapolation is rarely need for more than $\approx5M_0$.

Note that when the AH finder fails and either a less accurate or 
extrapolated shape is used,
this shape only guides the position and orientation of
the excision surface, {\em not} its size. This is important
for stability and to prevent the AH robustness problems from 
adversely affecting the exterior solution, as typically the
approximate AH grows with time, possibly even moving outside
the event horizon.
However, as the AH shape is used to measure the mass and
angular momentum of the black holes the corresponding plots
(see Sec.~\ref{sec_cdg} and Sec.~\ref{sec_bbh}) {\em do} reflect the 
problems of the AH finder.

\section{Case Studies of the Effect of Constraint Damping and Choice of Gauge}\label{sec_cdg}

In this section some results are given on the effect of constraint damping
and choice of source function evolution in a dynamical simulation. Little
is known about these two topics in general, namely, whether constraint
damping will work in generic $3+1$ evolutions to suppress the growth
of constraint violating modes, or what classes of source function evolution
equations could be used to achieve various well-behaved slicings of dynamically
evolved spacetimes. These questions certainly cannot be answered with a few
cases studies, and that is not the intended purpose; rather, the material
presented here
is to demonstrate how constraint damping and the gauge evolution 
equation (\ref{t_gauge})
affects the evolution of the class of asymptotically flat, scalar field collapse
binary black hole spacetimes considered here.

The majority of results in this section will be from an equal mass head-on 
collision,
though it will be shown that qualitatively similar conclusions apply to 
the orbital scenarios described in Sec.~\ref{sec_bbh}. The reason for looking at 
a head-on collision is that the simulation can be run in axisymmetry, and
so consumes less of the limited computer resources that are 
available\footnote{for example, a typical axisymmetric calculation
takes on the order of an hour to several days, running on 
between 4 and 16 Pentium IV class nodes of a Beowulf cluster. The 3D calculations
take on the order of several days to a month to complete, running on
between 32 and 128 Pentium IV class nodes}.
This makes it practical to do a more thorough survey of the
effects of different constraint damping and gauge evolution parameters.

The parameters for the head-on collision, including initial separation
and simulation parameters (grid hierarchies, dissipation, etc.) were chosen
to be close to the parameters used in the 3D simulations.
Units are scaled to $M_0=2m_0$,
twice the initial mass of one of the black holes in
the binary as measured by the area of its apparent 
horizon\footnote{in~\cite{paper2} the units were scaled by $m_0$ instead
of $2m_0$}. One common
convention in the literature is to scale by the ADM mass of the
spacetime; that is not used here because it includes the part of the
scalar field that does not fall into the black holes, which could
contribute as much as $10-15\%$ to the total mass of the spacetime. 
The initial conditions for the scalar field
in the the head-on collision simulation are as follows. The scalar field
pulses are initially at rest (i.e. zero boost), and placed such
that the initial coordinate separation 
of the two coordinate centers of the AH's that are first detected (at
$t\approx 2.5M_0$) is $8.90M_0${\ }\footnote{the coordinate centers
in unscaled units are at $x=\pm 0.08$, where $x$ runs along the axis
of symmetry. The tangent compactification function used
maps $x=\pm 1$ to spatial infinity, thus $x=\pm 0.08$ is well
within the linear range of this transformation~\cite{paper1}}; the initial proper 
separation measured along a 
coordinate line connecting the centers of the AH's from the
surface of one AH to the next is $10.8M_0$.
With these initial conditions the black holes merge at $t\approx 37M_0$,
and the energy emitted in gravitational waves is approximately 
$0.0010 M_0$, calculated using (\ref{Ep}) at a radius $r=50M_0$.
The ``canonical'' simulation relative to which others will be compared
uses a value of the constraint damping parameter $\kappa=1.80/M_0$ (\ref{efe_h_cd}), 
and gauge evolution parameters $\xi_1=0.50/M_0^2$, $\xi_2=5.4/M_0$ and
$\eta=1$ (\ref{t_gauge}). Factors of $1/M_0$ have been inserted
according to the dimensionality of the terms the constants
multiply in the equations. There is no particular reason why these specific
numbers were chosen, and, as 
will be demonstrated below,
no fine-tuning of the parameters is necessary---the only 
requirement is that in magnitude the constants be
of order unity relative to the smallest relevant length scale
in the problem ($M_0$) to produce a noticeable effect.

\subsection{Convergence}\label{sec_cnv}

\begin{figure}
\begin{center}
\includegraphics[width=8.5cm,clip=false]{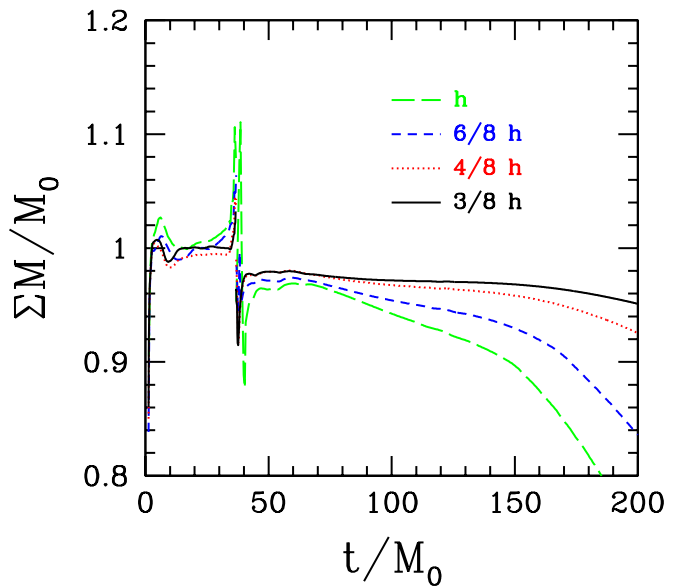}
\end{center}
\caption{
The normalized sum of total AH mass (\ref{smarr}) as a function
of time, for the head-on collision simulations described in
Sec.~\ref{sec_cdg} {\em with} constraint damping (compare Fig.\ref{fig_M_thc_wo_cd}). 
This plot demonstrates convergence to
a conserved mass before and after the merger. The ``spiky''
behavior about the merger point (around $t=37M_0$) is 
due to AH finder problems as discussed in Sec.\ref{sec_eah}.
}
\label{fig_M_thc_w_cd}
\end{figure}

For convergence tests four characteristic grid resolutions are used, 
summerized in Table~\ref{tab_res} below.
In axisymmetry computational resources
{\em are} available to go to higher resolution,
though this has not been done to facilitate comparison 
with the 3D simulations, which to date have only been run with resolutions
comparable to the three lower resolutions in Table~\ref{tab_res}.
The first measure of the 
accuracy/convergence properties of the solution is shown 
in Fig.~\ref{fig_M_thc_w_cd}, which plots the sum of AH masses of
black holes as a function of time. Assuming the AH is a good approximation
to the event horizon, the sum of AH masses should approach a 
constant after scalar field accretion ends, and at late times
after the merger. As seen in the figure,
as resolution increases the conservation of AH mass improves where it
is expected to do so.
Around the time of the merger there is a sharp
spike-like feature in this function. This is a reflection of
the robustness problems in the AH finder as discussed in Sec.~\ref{sec_eah};
however, the amount of time that this anomalous behavior
persists does seem to converge away.

\begin{figure}
\begin{center}
\includegraphics[width=8.5cm,clip=false]{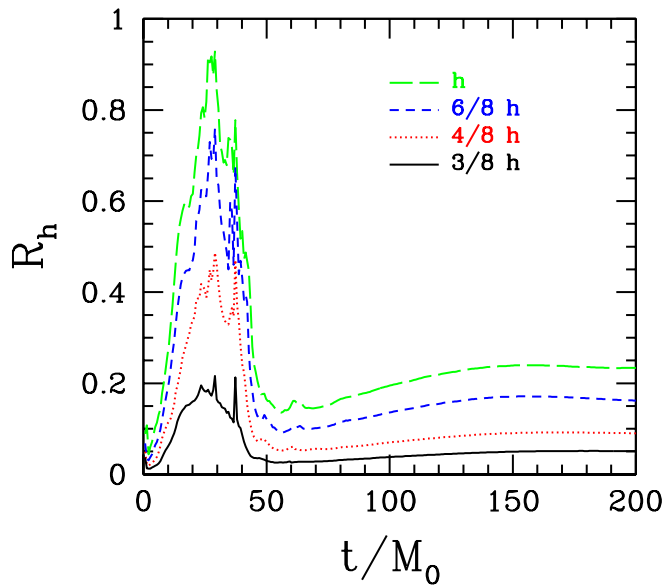}
\end{center}
\caption{
A norm of the residual of the Einstein equations (\ref{efe}), calculated
as discussed in Sec.~\ref{sec_cnv}, for the head-on collision 
simulations described in Sec.~\ref{sec_cdg} {\em with} constraint damping
(compare Fig.\ref{fig_efe_thc_wo_cd}, though note the different vertical
and horizontal scales).
This demonstrates the convergence of the solution---see also Fig.~\ref{fig_cf_efe_thc_w_cd}.
}
\label{fig_efe_thc_w_cd}
\end{figure}

Fig.~\ref{fig_efe_thc_w_cd} shows $\mathcal{R}_h$, an $\ell_2$ norm of a residual
of the Einstein equations in original form(\ref{efe}).
Note that by monitoring this particular residual rather 
than only the constraints (\ref{h_const}), or the
residual of the equations in GH form (\ref{efe_h_cd}), one has
an additional check that errors have not been introduced 
in going from (\ref{efe}) to (\ref{efe_h_cd}), and that the
requirements for solutions of the GH form of the equations to
be solutions of the Einstein equations are satisfied.
$\mathcal{R}_h$ is computed as follows.
First, the residual $\mathcal{R}_{ab}(i,j,k)_h$
of all ten field equations
\be
R_{ab}-8\pi\left(T_{ab}-\frac{1}{2}g_{ab} T\right)
\ee
is calculated using standard second order accurate finite
difference approximations
from a numerical solution obtained with a characteristic discretization scale
$h$, 
at a given grid location $(i,j,k)$ (or ($i,j$) in axisymmetry).
The residual at each point is normalized 
by the $\ell_2$ norm of all second derivatives of all
metric elements at the same point. This somewhat arbitrary normalization is simply
to give a convenient scale to plot the residual; the numerical
value of the residual from {\em one} simulation is not particularly meaningful,
rather it is the convergence to zero of the residual with increasing
resolution that is a test of the correctness of the solution scheme. 
After computing the normalized $\mathcal{R}_{ab}(i,j,k)_h$,
the infinity norm over the ten residuals is taken to define the 
residual $\mathcal{R}(i,j,k)_h$
of the grid location $(i,j,k)$. The quantity $\mathcal{R}_h$, as shown
in Fig.~\ref{fig_efe_thc_w_cd}, is then the $\ell_2$ norm of 
$\mathcal{R}(i,j,k)_h$ over the domain, though for 
computational convenience the points included in the norm
were restricted to a
uniform distribution of size $129^3$ ($129 \times 65$ in axisymmetry)
that encompassed roughly $50 M_0^3$ of the domain centered about
the origin. The residual outside this region drops to zero quite
rapidly. At early times (the first $\approx 40 M_0$) the residual is dominated 
by the scalar field---i.e. it is largest in the vicinity
of the outgoing waves of scalar field that did not fall into the black 
holes---which is the reason for the relatively large values of the residual
then.

For a finite difference numerical solution that is in the convergent 
regime one expects any quantity 
$Q_h(t)$, calculated from a solution obtained with a
discretization scale $h$, to have a Richardson expansion of the form
\be
Q_h(t) = Q(t) + e_{Q1}(t) h^n + e_{Q2}(t) h^{2n} + ...,\label{rich_exp}
\ee
where $Q(t)$ is the continuum value, $e_{Q1}(t),e_{Q2}(t),...$ are a set of error 
functions that depend on $Q(t)$ though not the
resolution, and $n$ is the order of convergence of the numerical 
technique\footnote{In an adaptive solution scheme one does not 
have a single discretization scale $h$,
nevertheless one would still expect a similar expansion with some
characteristic scale $h$ describing how well features of the solution
are resolved}. With second order accurate discretization $n=2$.
For the residual $\mathcal{R}_h(t)$, the continuum value $\mathcal{R}(t)=0$,
and so ignoring higher order terms one has
\be
\mathcal{R}_h(t) = e_{R1}(t) h^n 
\ee
Using residuals from simulations with two resolutions $h_1$
and $h_2$ one can eliminate the unknown error term from the
above equation, and solve for $n$:
\be\label{n_h1h2}
n(h_1,h_2)(t) = \frac{\log\left(\mathcal{R}_{h_1}(t)/\mathcal{R}_{h_2}(t)\right)}{\log(h_1/h_2)}.
\ee
A measurement of $n(h_1,h_2)(t)$ can be used as a convergence test:
$n(h_1,h_2)$ tending toward $2$ as resolution increases implies
that the assumed expansion in (\ref{rich_exp}) is valid and the
ignored higher order terms are small.
If the continuum value $Q(t)$ is not known,
one can estimate the error in $Q_{h2}(t)$ calculated from a simulation with mesh
spacing $h_2$ if a second simulation with spacing $h_1$ is available,
{\em and} assuming both simulations are in the convergent regime:
\be\label{rich_err}
Q(t)-Q_{h_2}(t)\approx e_{Q_1}(t)h_2^n \approx h_2^n\frac{Q_{h_2}(t)-Q_{h_1}(t)}{h_2^n-h_1^n}.
\ee

Fig.~\ref{fig_cf_efe_thc_w_cd} below shows
$n$ calculated for the sequence of residuals shown in Fig.~\ref{fig_efe_thc_w_cd}.
As the resolution of the pair of mesh spacings $(h_1,h_2)$ increases,
one expects the numerically calculated
$n$ to tend to 2; this trend is evident
in the figure, especially at later times.
At early times this behavior
is not as apparent, though again due to the scalar field. Relatively speaking, the
scalar field is under-resolved compared to the smallest scale of 
interest (the black holes): the smallest length scale
in the scalar field is $\Delta$ (\ref{sf_id}), which was
chosen to be roughly $M_0/2$, whereas the corresponding
length scale for a black hole is its diameter, which initially is
$4 m_0 = 2 M_0$. For these simulations the scalar field is merely a vehicle
to produce black holes, so that it is somewhat under-resolved
is not a source of much concern.

\begin{table}
{\small
\begin{tabular}[t]{| l l || c | c | c |}
\hline
 & ``Resolution'' & wave-zone res.& orbital-zone res.& BH res.\\
\hline
\hline
 & h     & $1.7  M_0$ & $0.23 M_0$  & $ 0.057 M_0$ \\
 & 6/8 h & $1.3  M_0$ & $0.17 M_0$  & $ 0.043 M_0$ \\
 & 4/8 h & $0.85 M_0$ & $0.12 M_0$  & $ 0.029 M_0$ \\
 & 3/8 h & $0.64 M_0$ & $0.087 M_0$ & $ 0.021 M_0$ \\
\hline
\end{tabular}
}
\caption{The four sets of characteristic resolutions used in
simulation results presented here, were each resolution 
is labeled relative to the coarsest resolution $h$ (only the
three lower resolutions have been used for non head-on mergers). 
The grid is adaptive with a total of 8 levels of refinement,
and the coordinate system is compactified. The wave zone
is defined to be at $r=50 M_0$, the orbital zone within about $r=10 M_0$
and the black hole zone is within $2-3M_0$ of each apparent horizon.
A CFL (Courant-Friedrichs-Lewy) factor
of $0.15$ was used in all cases.
}
\label{tab_res}
\end{table}

\begin{figure}
\begin{center}
\includegraphics[width=8.5cm,clip=false]{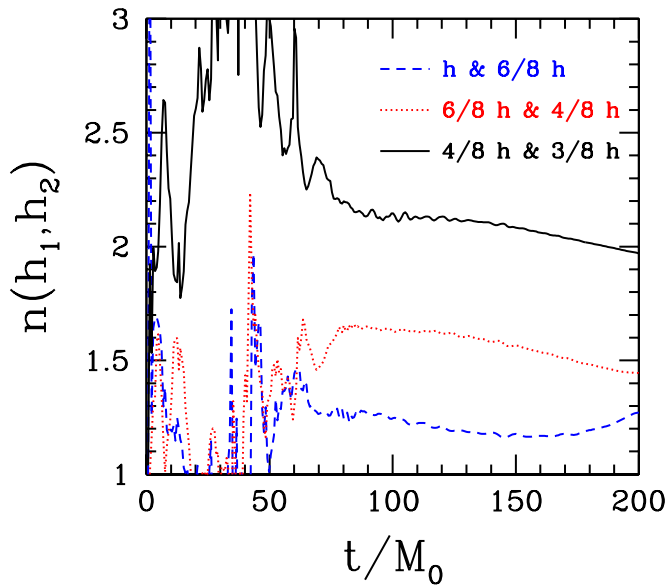}
\end{center}
\caption{
The order of convergence (\ref{n_h1h2}) calculated from the head-on collision
simulations described in Sec.~\ref{sec_cdg} (with constraint damping),
and using the residual data shown in Fig.~\ref{fig_efe_thc_w_cd}. The
discretization scheme is second order, and so one would expect
$n(h_1,h_2)$ to asymptote to 2 as the resolution is increased.
This is evident in the figure, particularly
after most of the scalar field has left the domain (around $t\approx 40-50M_0$).
}
\label{fig_cf_efe_thc_w_cd}
\end{figure}

\subsection{Constraint Damping}

\begin{figure}
\begin{center}
\includegraphics[width=8.5cm,clip=false]{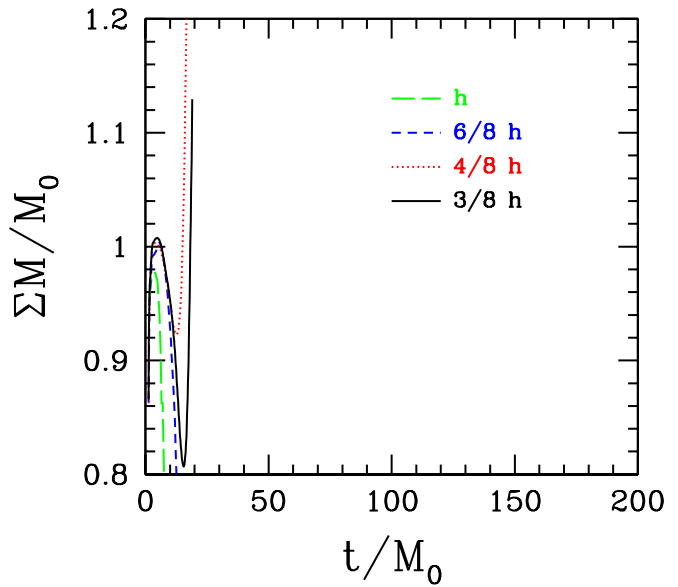}
\end{center}
\caption{
The normalized sum of total AH mass (\ref{smarr}) as a function
of time, for the head-on collision simulations described in
Sec.~\ref{sec_cdg} and {\em without} constraint damping;
compare Fig.~\ref{fig_M_thc_w_cd}. The curves end when the simulations
crashed.
}
\label{fig_M_thc_wo_cd}
\end{figure}

\begin{figure}
\begin{center}
\includegraphics[width=8.5cm,clip=false]{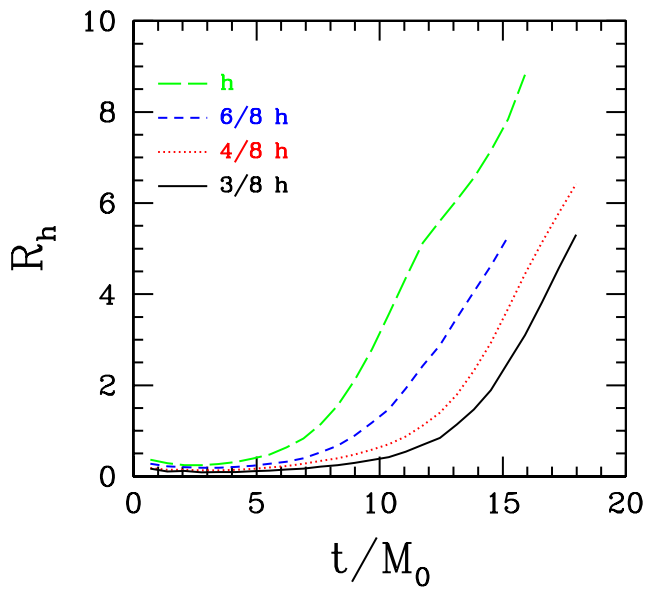}
\end{center}
\caption{
A norm of the residual of the Einstein equations (\ref{efe}), calculated
as discussed in Sec.~\ref{sec_cnv}, for the head-on collision 
simulations described in Sec.~\ref{sec_cdg} {\em without} constraint damping;
compare Fig.\ref{fig_efe_thc_w_cd}, though note the different vertical
and horizontal scales. Again, convergence is evident, though the rapid
growth of the residual with time prevents useful results from being obtained
at modest resolution.
}
\label{fig_efe_thc_wo_cd}
\end{figure}

\begin{figure}
\begin{center}
\includegraphics[width=8.5cm,clip=false]{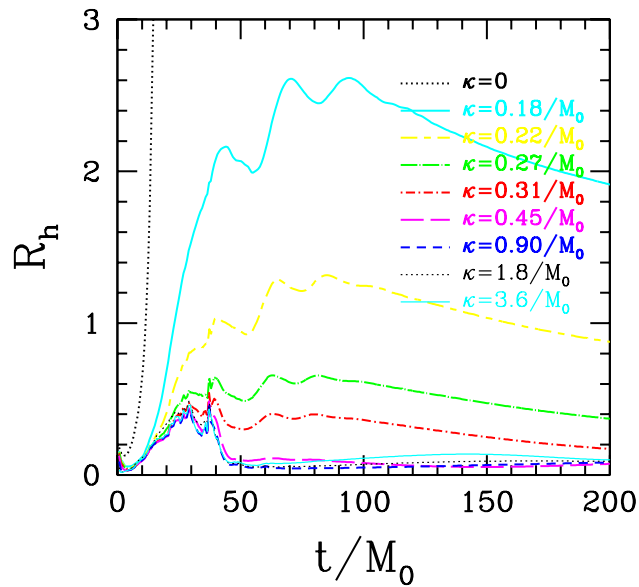}
\end{center}
\caption{
Norms of the residual of the Einstein equations (\ref{efe}), calculated
as discussed in Sec.~\ref{sec_cnv}, for a set of identical ``$4h/8$'' resolution
head-on collision simulations, though with differing values of the 
constraint damping parameter $\kappa$. The results shown in
Fig.~\ref{fig_efe_thc_w_cd} were 
obtained with $\kappa=1.8/M_0$ (though note the different vertical scale). 
}
\label{fig_efe_thc_cd_comp}
\end{figure}

This section contains a few results demonstrating the effectiveness of constraint
damping. Fig.~\ref{fig_M_thc_wo_cd} is the equivalent of Fig.~\ref{fig_M_thc_w_cd},
showing the sum of apparent horizon masses as a function of time,
though now the corresponding set of simulations have been run with
the constraint damping parameter $\kappa=0$. All of the simulations without 
constraint damping ``crashed'' before the holes merged, which is why
the curves end abruptly. Fig.~\ref{fig_efe_thc_wo_cd} shows plots of the
residual $\mathcal{R}_h$ for the simulations without constraint damping---compare
to Fig.~\ref{fig_efe_thc_w_cd}, though note the different scales used in
the two plots. This clearly shows how effective constraint damping is,
though does not tell us how large $\kappa$ needs to be to start to work.
To get an idea for what the required magnitude of $\kappa$ is,
a set of $4/8 h$ resolution simulations were run with a range
of different $\kappa$'s---Fig.~\ref{fig_efe_thc_cd_comp} shows the 
residual from these simulations. This demonstrates
that $\kappa$ needs to be of order unity (relative to $1/M_0$), though
no fine tuning is necessary, i.e. there are a range of values $\kappa>\kappa_0$
that are all essentially equivalent in damping rapid growth in the constraints.

\subsection{Source Function Evolution}

\begin{figure}
\begin{center}
\includegraphics[width=8.5cm,clip=false]{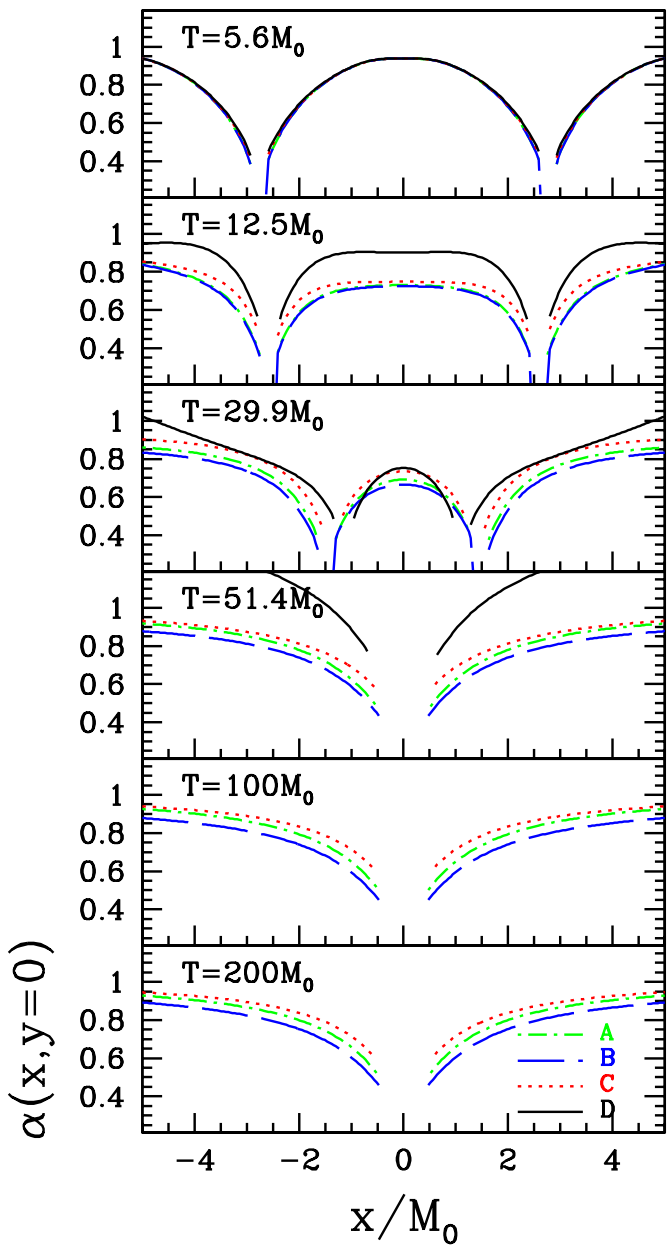}
\end{center}
\caption{
The lapse function $\alpha$ along the axis of symmetry recorded at several times
from a set of ``$4h/8$'' head-on collision simulations as described in
Sec.~\ref{sec_cdg}, with different parameters in the  
evolution equation (\ref{t_gauge}) used for $H_t$---see Table~\ref{tab_slice_def}.
Case A features the same values
as the simulation results shown in Fig.'s 
\ref{fig_M_thc_w_cd}-\ref{fig_efe_thc_cd_comp}, case B
is pure harmonic gauge, case C has a larger $\eta$ coefficient
than case A and so is more effective at keeping the lapse
closer to $1$, and case D is the same as A though with zero  
damping coefficient $\xi_2$. Without the damping term the
driving term, controlled by $\xi_1$, is too large in this example, and $\alpha$
overshoots $1$ by a sufficiently large factor that the simulation
crashes, which is why curve $D$ is not present
in the last two frames.
}
\label{fig_alpha_slicing}
\end{figure}

Fig.~\ref{fig_alpha_slicing} demonstrates the effect of changing
parameters in the source function evolution equation (\ref{t_gauge})
for the head-on collision simulation; Table~\ref{tab_slice_def} shows
the values of the parameters for each set of curves plotted. To avoid
clutter in the figure results are given from only 4 different simulations,
however these are quite representative of the range of behavior seen with
this gauge evolution equation.
The figure shows the lapse function $\alpha$ along the axis of symmetry
at a few select times (at $t=0$, $\alpha\approx 1$ and is identical in
all cases). There are several features shown in the figure
worth mentioning. First, the curve ($D$) demonstrates source function evolution
without a damping term; this tends to cause oscillations
about $1$ in the lapse, and with sufficiently large values of $\xi_1$
the oscillations grow to the point where the
code crashes. This is the reason the ($D$) curve is not visible
at late times. 

Second, in cases with a source term ($A$,$C$,$D$) 
$\alpha$ does not evolve to $1$ in the strong-field regime even at late times, though is
closer to $1$ than harmonic gauge ($B$). For example, near merger
at $t=29.9M_0$ the minimum value $\alpha_{\rm{min}}$ outside the excision
surface for harmonic gauge is 
$\alpha_{\rm{min}}=0.213$ compared to $0.505$ for case $C$,
and at $t=200M$ $\alpha_{\rm{min}}=0.458$ (harmonic) versus
$0.607$ ($C$). These differences might not seem too significant,
and at least for head-on collisions all cases except $D$ 
work adequately.
With non head-on collisions where there is significant orbital
motion prior to merger, the small
values that the lapse tends to with harmonic gauge seem to be problematic in that instabilities
form before a common horizon is detected. There is not convincing
evidence that this is coordinate problem as opposed to a numerical issue, though
given how ``expensive'' simulations are in $3D$ and that the non-harmonic
gauge works, this issue has not yet been explored in any detail.

Third, notice that for all gauge evolution schemes that are long-time
stable ($A,B,C$), at late times the solution approaches
a largely time independent form. This gauge condition therefore
seems to be ``symmetry-seeking'', at least for this
class of spacetimes.

\begin{table}
\begin{tabular}[t]{| l l || c | c | c |}
\hline
 & label & $\xi_1$ & $\xi_2$ & $\eta$ \\
\hline
\hline
 & A & $0.5/M_0^2$ & $5.4/M_0$  & $1$ \\
 & B & $0        $ & $0$        & $0$ \\
 & C & $0.5/M_0^2$ & $5.4/M_0$  & $7$ \\
 & D & $0.5/M_0^2$ & $0$        & $1$ \\
\hline
\end{tabular}
\caption{
Values of the parameters in the gauge evolution equation
(\ref{t_gauge}) used for the comparison simulation results
shown in Fig.~\ref{fig_alpha_slicing}.
}
\label{tab_slice_def}
\end{table}

\subsection{Comparison with 3D evolutions}

\begin{figure}
\begin{center}
\includegraphics[width=8.5cm,clip=false]{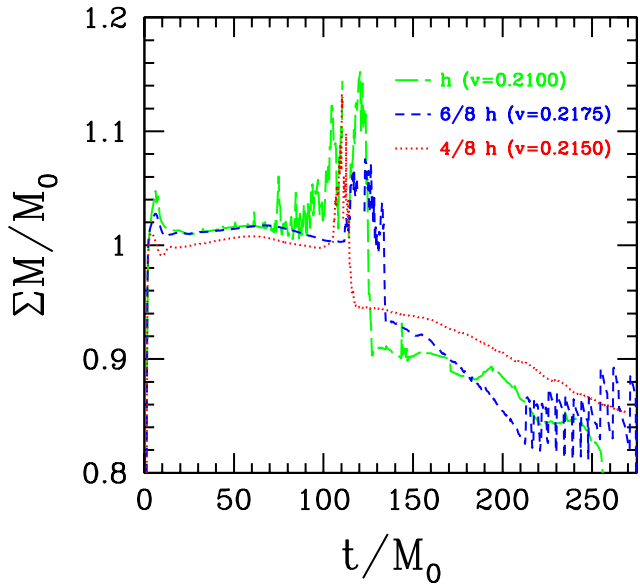}
\end{center}
\caption{
The normalized sum of total AH mass (\ref{smarr}) as a function
of time for 3 merger simulations (see Fig.\ref{fig_d_one_half}), as 
described in Sec.~\ref{sec_bbh}.
In each case the boost parameter was chosen so that roughly 
one and a half orbits is completed before merger, to better facilitate
the comparison of the effect of resolution on solution accuracy
(see the discussion in Sec.~\ref{sec_bbh}). The late-time
problems with the AH finder for the $6h/8$ case
seems to set in when numerical error causes the angular momentum
of the black hole to approach extremality--see Fig.~\ref{fig_a_one_half}.
}
\label{fig_M_one_half}
\end{figure}

\begin{figure}
\begin{center}
\includegraphics[width=8.5cm,clip=false]{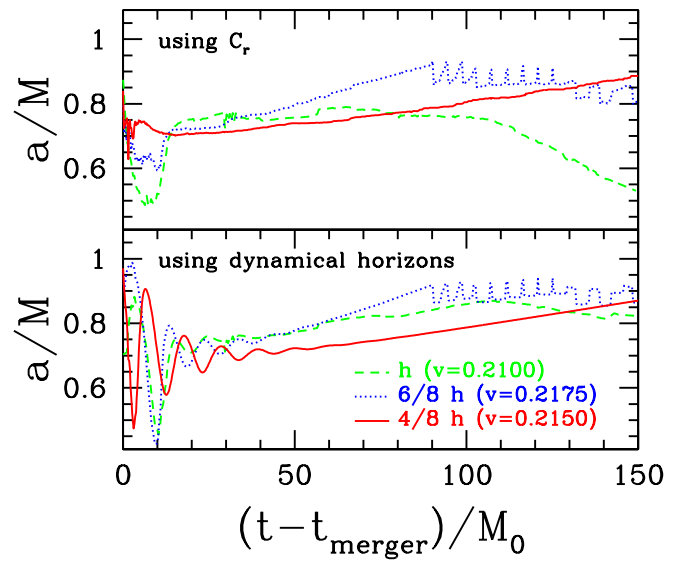}
\end{center}
\caption{
Estimates of the angular momentum of the final black hole,
corresponding to the simulations shown in Fig.'s~\ref{fig_M_one_half} 
and \ref{fig_d_one_half}, calculated using (\ref{a_cr}) and
the dynamical horizons framework in the top and bottom plots
respectively. To facilitate the comparison the time has been
shifted so that $t=0$ corresponds
to the moment an encompassing AH is first detected.
Note that with increased resolution the drift
in $a/M$ decreases at late times, as to be expected
in a convergent solution.
The $6h/8$ resolution case demonstrates what is seemingly 
a further source of robustness problems in the AH finder when
the angular momentum approaches extremality (the 
spin-up is numerical-error driven here).
}
\label{fig_a_one_half}
\end{figure}

Here it is demonstrated that some of the results just discussed
for the 2D, axisymmetric head-on collisions apply to the full
3D merger problems described in Sec.~\ref{sec_bbh}, by comparing a select case.
Fig.~\ref{fig_efe_2d_3d_comp} below shows the residual $\mathcal{R}_h$ 
from three 3D simulations with similar
grid structure to the three lowest resolution 2D results 
shown in Fig.~\ref{fig_efe_thc_w_cd}; the sum of AH masses with
time is shown in Fig.~\ref{fig_M_one_half}; and though there is no
direct comparison to the 2D results Fig.~\ref{fig_a_one_half} shows
the estimated Kerr spin parameter $a$ after the merger. The parameters for the 3D simulations
were chosen so that the equal mass merger occurs in roughly an orbit and a
half for each resolution---see Fig.~\ref{fig_d_one_half}.
Note that the boost
parameters differ slightly for each resolution; this, as discussed in Sec.\ref{sec_bbh},
is due to the sensitivity of the resultant orbit to the initial data parameters.
Of course, in the continuum limit one would expect convergence to a definite
orbit for a given boost parameter, or conversely, given a desired orbit
the required boost parameter should 
converge to a definite value. However, what Fig.~\ref{fig_efe_2d_3d_comp} is
meant to show is the behavior of the residual $\mathcal{R}_h$ 
with time and as resolution varies, and so for a meaningful comparison
simulations with similar grid structures and run-times were used.

\begin{figure}
\begin{center}
\includegraphics[width=8.5cm,clip=false]{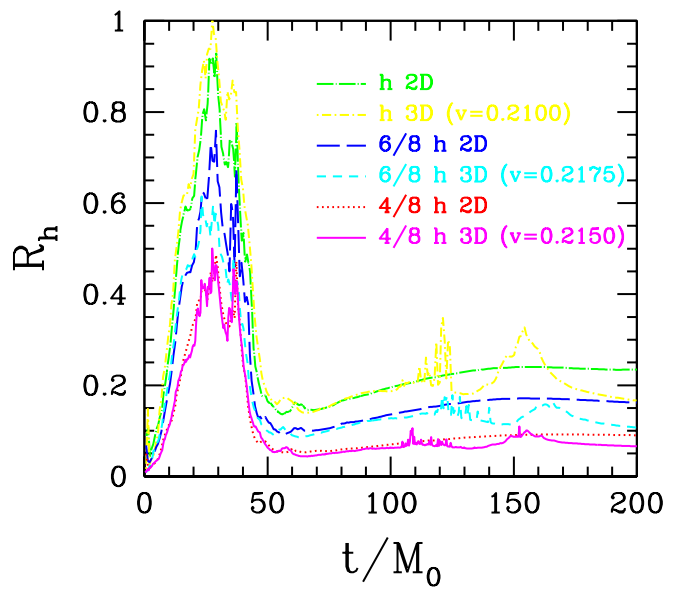}
\end{center}
\caption{
Comparison plots of the norm of the residual of the Einstein 
equations (\ref{efe}), calculated
as discussed in Sec.~\ref{sec_cnv}, between the head-on collision 
simulations described in Sec.~\ref{sec_cdg} (labeled $2D$---see 
Fig.\ref{fig_efe_thc_w_cd}) and
non head-on mergers (labeled $3D$) as discussed in Sec.~\ref{sec_bbh}.
$3h/8$ resolution simulations were not run in 3D due to lack of computational resources. 
Note also that the momentary increase in the residual near
$t=100M_0-150M_0$ in the $3D$ simulations is around the time
of merger.
}
\label{fig_efe_2d_3d_comp}
\end{figure}

\begin{figure}[b]
\begin{center}
\includegraphics[width=8.5cm,clip=false]{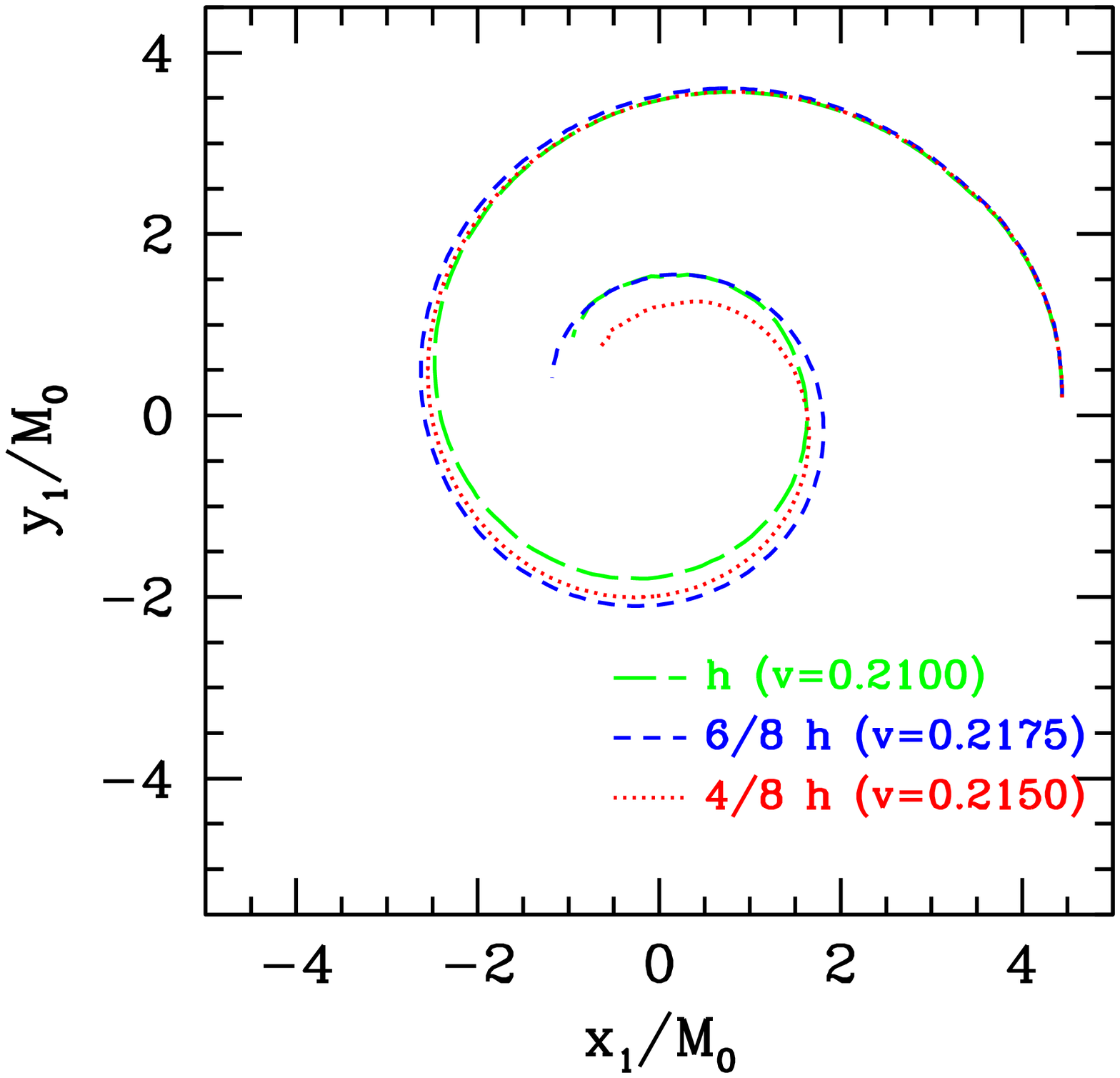}
\end{center}
\caption{
A plot of the orbits of the $3D$ simulations used to compare
with the head-on collision results (Fig.~\ref{fig_efe_2d_3d_comp}).
What is shown for each case is the coordinate position of the
center of the apparent horizon of one of the black holes.
}
\label{fig_d_one_half}
\end{figure}

\section{Scalar Field Collapse Driven Binary Black Hole Spacetimes}\label{sec_bbh}

In this section some results are presented from an ongoing study of scalar
field collapse driven black hole merger simulations. The disadvantages
of this approach to constructing an orbit, from the point of view
of simulating astrophysically relevant binary configurations, is that
one does not have {\em a priori} control over the orbit
that will result, and perhaps more problematic is that in the
regime where the evolution is started the radiation-reaction is sufficiently
strong that it is not easy to deduce an astrophysical
orbit that the binary might have evolved from. Therefore, these binary
might be in a region of parameter space that are {\em not}
shared by astrophysical systems. However, from a theoretical
perspective of trying to understand the dynamics of binary black hole
coalescence in general relativity, this scalar field collapse initial
data is a straight-forward and simple way to probe a large and interesting
region of parameter space.

Here the focus will be on one particular family of initial conditions:
identical initial scalar field distributions ($A=0.3$ and $\Delta=0.827M_0$)
except that one
is located at $(x,y,z)=(4.45M_0,0,0)$ and given a boost with boost parameter
$v$ in the positive-$y$ direction, while the other is located at 
$(x,y,z)=(-4.45M_0,0,0)$ and given a boost $v$ in the negative-$y$
direction---see (\ref{sf_id}) and the discussion afterwards.
Approximately $85\%$ of the energy
in the scalar field falls into the black holes that form, while
the remaining energy radiates away on roughly an orbital
light-crossing time scale. 

\begin{figure}
\begin{center}
\includegraphics[width=8.5cm,clip=false]{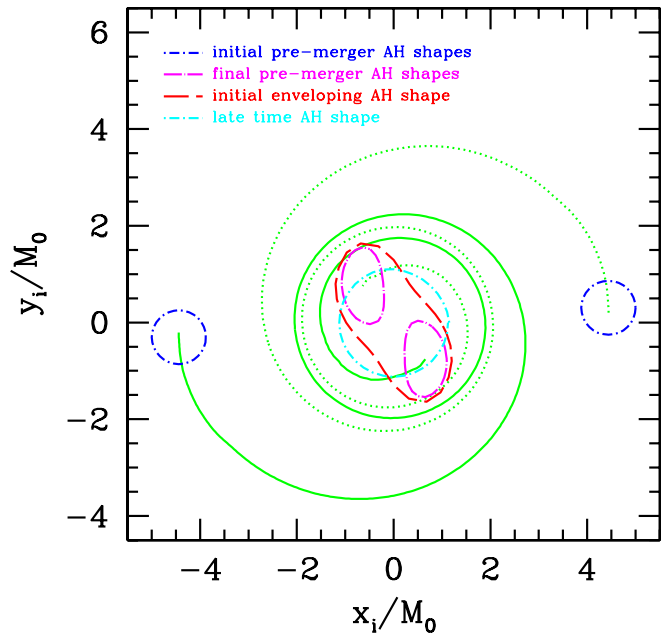}
\end{center}
\caption{
A plot of the orbit of the $v=0.21906$, $6h/8$-resolution merger
simulation. The spiral trajectories are 
the coordinate positions of the
centers of the apparent horizons of the two black holes before merger,
and the labeled curves show the coordinate shapes of the AH's,
in the $z=0$ plane, at select times.
}
\label{fig_d_Lm3_t21w}
\end{figure}

\begin{figure}
\begin{center}
\includegraphics[width=8.5cm,clip=false]{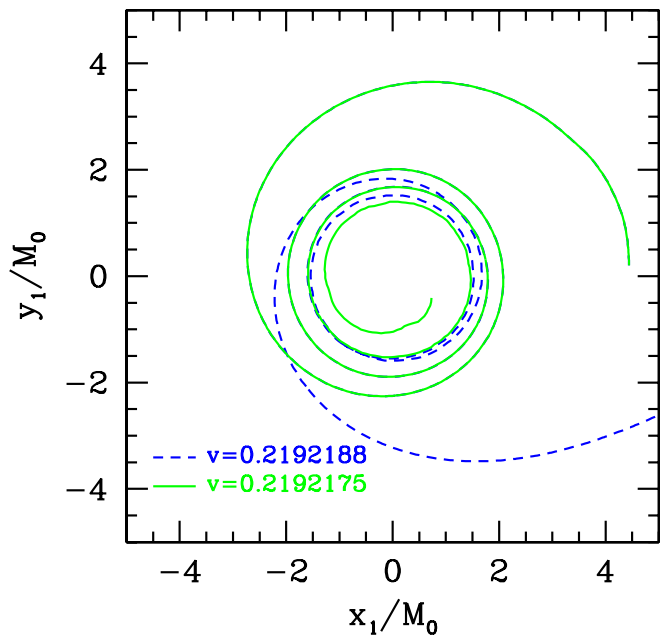}
\end{center}
\caption{ 
A plot of the orbits of two of the $6h/8$-resolution simulations
that have been tuned closest to the bifurcate-like point in boost-parameter
space (see Table~\ref{tab_sfc_data}). The merger case orbit ends
when an encompassing AH is detected.
}
\label{fig_d_Lm3}
\end{figure}

\begin{figure}[b]
\begin{center}
\includegraphics[width=8.5cm,clip=false]{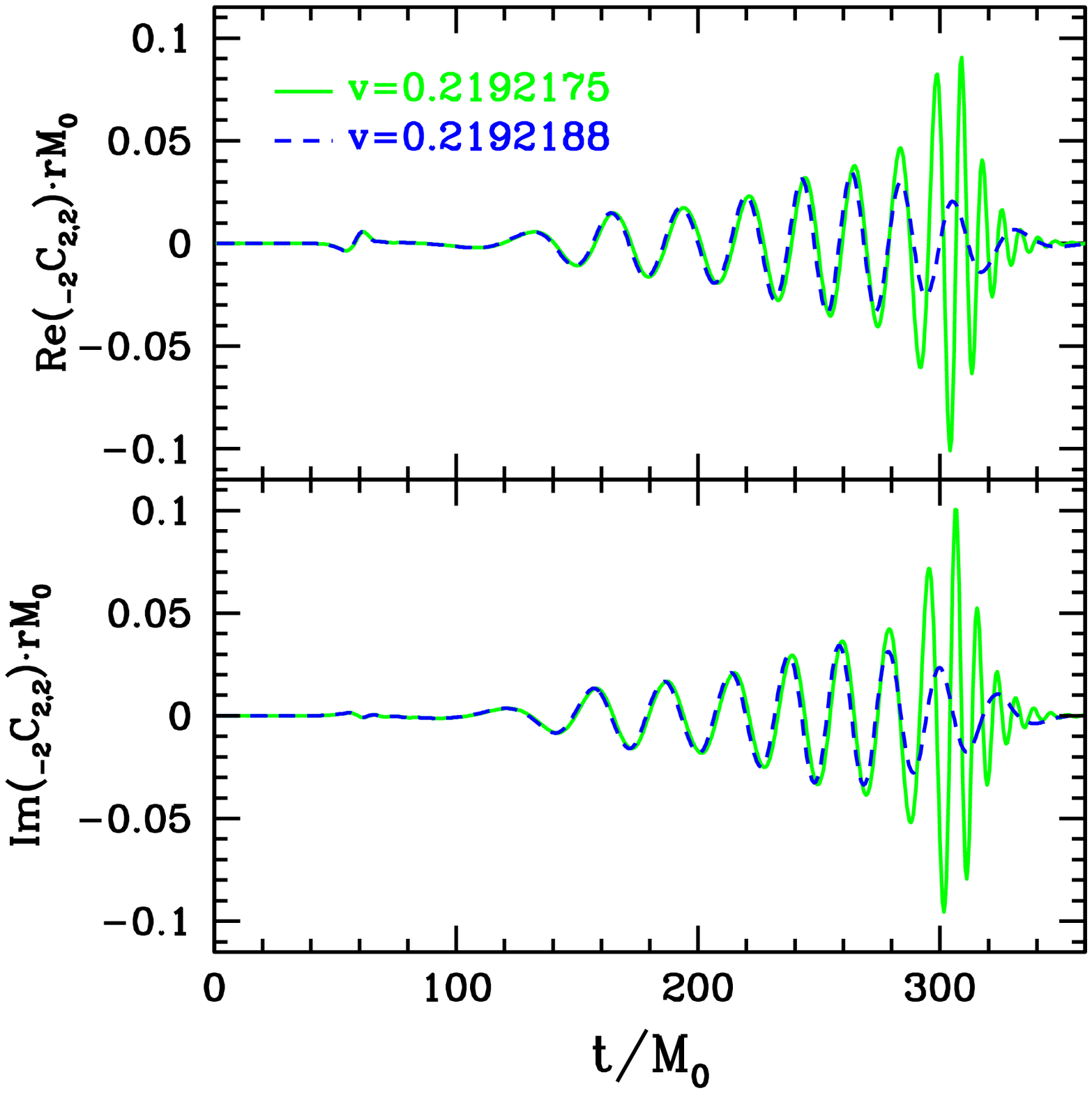}
\end{center}
\caption{ 
Gravitational waveforms from the two simulations
depicted in Fig.~\ref{fig_d_Lm3}. What is shown here 
are the two polarizations of 
one of the dominant spin weight $-2$ spherical harmonic
components of $\Psi_4$ (\ref{psi4_c_def}), calculated
over a coordinate sphere a distance $r=50M_0$ from the 
origin, and normalized by $rM_0$.
}
\label{fig_Lm3_psi4}
\end{figure}

This choice of initial data gives a single parameter $v$
that can be varied.
The black holes are initially sufficiently far 
apart---a coordinate (proper) separation of $8.90M_0$ ($10.8M_0$)---that 
with a large enough
boost parameter one would expect the system to be unbound.
At the other extreme, $v=0$ will
give a head on collision. Thus there is a range of values
$0\leq v_c$ for some $v=v_c$ that will result in a merger, and it is 
this region of parameter space we want to explore. For the purposes
of studying the gravitational radiation produced by mergers, and
in particular comparing the orbital to plunge and ringdown part
of the waves, it is useful to have parameters that result in several orbits
before the merger--see \ref{fig_d_Lm3_t21w} for a plot of the orbit 
from one such parameter. This regime is where these BBH simulations become 
quite interesting, in that there is apparently
extreme sensitivity of the solution as a function of the boost parameter.
In particular, the dependence of the number of orbits $n$ as a function
of $v$, $n(v)$ is quite reminiscent of certain aspects of
critical phenomena in gravitational collapse
in that there are apparently
two distinct end states, merger of the black holes for $v<v^\star$ 
versus separation for $v>v^\star$, and
the closer $v$ is to $v^\star$ the more orbits before merger/separation
that are observed
---see Table~\ref{tab_sfc_data} below for data from several
simulations, Fig.~\ref{fig_d_Lm3} for a plot of the
orbits from two of the $6/8 h$ resolution runs that have been fine-tuned 
the most to date, and Fig.~\ref{fig_Lm3_psi4} for graphs of the
corresponding gravitational waves emitted\footnote{note however
that {\em no} claims are being made that this is {\em critical phenomena}
in any traditional sense of the word; rather, some of the
terminology is useful in describing these tentative results,
as is the bisection search strategy used in critical collapse studies
to find the threshold parameter $v^\star$}.
This behavior is also quite reminiscent of ``zoom-whirl'' orbits seen in the trajectories
of point-particle orbits about black holes\cite{zoom_whirl}, and might 
simply be the fully non-linear equivalent of the zoom-whirl phenomena.
In fact, for zoom-whirl particle orbits there {\em is} exponential
sensitivity of the number of orbits $N_w$ completed in each whirl phase
to the initial conditions : $p-p_s=e^{-a N_w}$, where $p$ is the semi-latus
rectum of the orbit, $p_s$ is the separatrix of bound orbits, and $a$
is some constant that depends (amongst other things) on the eccentricity
of the orbit and spin of the central black hole\cite{zoom_whirl_b}. Of course,
in the full solution where energy is lost through gravitational wave
emission $N_w$ can not be arbitrarily large. Furthermore,
$v^\star$ loosely defined above cannot be a ``true'' boundary between merger and separation, 
for one would still expect the orbits to be bound for $v$ slightly
greater than $v^*$---the binary will just ``whirl'' out some distance
before later merging.

Note that the sensitivity to initial conditions complicates
convergence testing (i.e. verification) of this interesting indication
that zoom-whirl like behavior is present
in equal mass binary merger spacetimes. This is because 
sensitivity to initial conditions implies that the ``bifurcate'' regime
of parameter space must be found {\em independently} 
 using {\em several} resolutions; i.e.
one cannot do a ``quick'' low-resolution search to find
$v\approx v^\star$, then perform a high resolution simulation to 
verify the low-resolution result, as each resolution will have
a different $v^\star$. Furthermore, the numerical error that accumulates
per orbit, as judged using AH mass conservation, is relatively
large, in particular for the lower resolution simulations,
where the AH mass changes by a few percent
per orbit. This is larger that the energy lost through gravitational
wave emission, and though
it does not exactly make sense to equate numerical error with 
physical energy, that these two ``effects'' are of the same
order of magnitude suggests one cannot yet draw significant conclusions
from these early results.

\begin{table*}[t]
\bigskip
{\bf h-resolution runs} \\
\begin{tabular}[t]{| l l || c | c | c | c | c | c | c |}
\hline
 & $v$ & $n$ & $p_{m}/M_0$ & $d_{m}/M_0$ & $m_f/M_0$ &  $a/m_f$ & $(E/M_0)$  \\
\hline
\hline
 & $0.21000$ & $1.3$ & $-$ & $-$ & $0.89\pm0.03$ & $0.75\pm0.05$ & $0.032$  \\ 
 & $0.21125$ & $1.4$ & $-$ & $-$ & $0.88\pm0.03$ & $0.74\pm0.05$ & $0.035$  \\ 
 & $0.21234$ & $2.3$ & $-$ & $-$ & $0.83\pm0.03$ & $0.73\pm0.05$ & $?$  \\ 
\hline
 & $0.21250$ & $2.7$  & $4.0$ & $3.6$ & $-$ & $-$ & $0.020$  \\ 
 & $0.21500$ & $1.5$  & $5.5$ & $4.6$ & $-$ & $-$ & $0.006$  \\ 
 & $0.22000$ & $1.0$  & $7.2$ & $5.8$ & $-$ &  $-$ & $0.005$  \\ 
\hline
\end{tabular} 

\bigskip
{\bf 6/8 h-resolution runs}\\
\begin{tabular}[t]{| l l || c | c | c | c | c | c | c |}
\hline
 & $v$ & $n$ & $p_{m}/M_0$ & $d_{m}/M_0$ & $m_f/M_0$ &  $a/m_f$ & $(E/M_0)$  \\
\hline
\hline
& $0.20960$ & $0.9$ & $-$ & $-$ & $0.97\pm0.01$ &  $0.65\pm0.03$ & $0.028$  \\ 
& $0.21750$ & $1.4$ & $-$ & $-$ & $0.92\pm0.01$ &  $0.72\pm0.03$ & $0.037$  \\ 
& $0.21875$ & $2.0$ & $-$ & $-$ & $0.88\pm0.01$ &  $0.70\pm0.03$ & $0.046$  \\ 
& $0.21906$ & $2.4$ & $-$ & $-$ & $0.86\pm0.01$ &  $0.70\pm0.03$ & $0.052$  \\ 
& $0.219180$ & $2.8$ & $-$ & $-$ & $0.82\pm0.02$ &  $0.70\pm0.05$ & $0.063$  \\ 
& $0.219200$ & $3.0$ & $-$ & $-$ & $0.80\pm0.02$ &  $0.75\pm0.05$ & $0.064$  \\ 
& $0.219209$ & $3.3$ & $-$ & $-$ & $0.78\pm0.02$ &  $0.71\pm0.05$ & $0.067$  \\ 
& $0.219214$ & $3.7$ & $-$ & $-$ & $0.75\pm0.02$ &  $0.71\pm0.05$ & $0.074$  \\ 
& $0.2192175$ & $3.9$ & $-$ & $-$ & $0.74\pm0.02$ &  $0.70\pm0.05$ & $0.076$  \\ 
& $0.2192181$ & $4.3$ & $-$ & $-$ & $0.73\pm0.02$ &  $0.71\pm0.05$ & $0.079$  \\ 
\hline
& $0.2192188$ & $4.9$ & $3.2$ & $3.0$ & $-$ &  $-$ & $0.058$  \\ 
& $0.21938$ & $2.5$ & $4.8$ & $4.2$ & $-$ &  $-$ & $0.019$  \\ 
& $0.22000$ & $1.9$ & $5.3$ & $4.4$ & $-$ &  $-$ & $0.014$  \\ 
\hline
\end{tabular}

\bigskip
{\bf 4/8 h-resolution runs}\\
\begin{tabular}[t]{| l l || c | c | c | c | c | c | c |}
\hline
 & $v$ & $n$ & $p_{m}/M_0$ & $d_{m}/M_0$ & $m_f/M_0$ &  $a/m_f$ & $(E/M_0)$  \\
\hline
\hline
 & $0.21500$ & $1.4$ & $-$ & $-$ & $0.945\pm0.005$ & $0.71\pm0.02$ & 0.042 \\ 
\hline
& $0.22000$ & $2.1$ & $5.7$ & $4.8$ & $-$ &  $-$ & $0.008$  \\ 
\hline
\end{tabular}

\caption{
Information extracted from the scalar field collapse driven
binary simulations run to-date, described in Sec.~\ref{sec_bbh}. Results from
three different resolutions (see Table~\ref{tab_res}) are 
shown (data from the $h$
resolution $v=0.21234$ run was lost, preventing the calculation of $E$
in that case).
$v$ is the boost parameter for the initial scalar field pulses and $n$ 
the number of orbits completed either before a merger,
or at the moment the binaries
reached the same coordinate separation as at $t=0$. For binaries that
did not merge, $p_m$ is the minimum proper separation measured at $t={\rm const.}$
along the coordinate line between, though {\em exterior to} the AH's, 
and $d_m$ is the minimum coordinate distance between the coordinate {\em centers}
of the AH's. For binaries that merged, the final mass and angular momentum of the
black holes, as estimated from AH properties, is listed. $E$ is the total
energy emitted in gravitational waves, calculated as described in Sec.~\ref{sec_solp}
at a coordinate radius of $50M_0$ from the origin. Note that the quoted
uncertainties in the final BH properties {\em only} include an estimated
uncertainty from looking at plots such as Fig.~\ref{fig_M_one_half} and 
\ref{fig_a_one_half}. To obtain an estimate of the numerical errors, 
one can use the Richardson expansion. For example by comparing the $n=1.4$ 
results between the $6/8h$ and $4/8h$ runs, and using (\ref{rich_err}),
the estimated errors in the $v=0.21500$ $4/8h$ simulation are:
$m_f=0.95\pm0.02M_0$, $a=0.71\pm0.01$ and $E=0.042\pm.004$. 
Once a larger set of simulations have been completed at higher resolution 
a more thorough analysis
of numerical errors will be performed.
}
\label{tab_sfc_data}
\end{table*}

\section{Conclusions}\label{sec_conclusion}
In this paper further details were given of the generalized
harmonic evolution scheme introduced in~\cite{paper1} and shown
to be capable of simulating binary black hole coalescence\cite{paper2}.
In particular, topics included were a demonstration of the effectiveness of constraint
damping in a fully non-linear setting, examples of the effect
of source function evolution on the time slicing of the spacetime,
a discussion of certain technical aspects of the code and problems with the robustness
of the apparent horizon finder, and
some results from an ongoing study of scalar field collapse driven
black hole binary simulations. There are many outstanding issues
that need to be explored, including the applicability of constraint
damping to more general scenarios and alternative evolution schemes,
gaining
more evidence for (or against) the zoom-whirl type behavior seen here,
studying a broader range of binary black hole merger initial conditions,
investigating a larger class of source function evolution equations 
(in particular to include non-trivial spatial source functions),
and beginning to extract some information from the non-perturbative
regime of the merger to aid in gravitational wave detection efforts.
The scalar field driven binaries are arguably not too useful in regards
to this latter point, primarily because of the difficulty in mapping
the resulting binary to (some approximation of) an astrophysical
binary. Nevertheless, general features of the waveforms could be
studied. Furthermore, what these simulations suggest is that to eventually obtain
accurate waveforms in the most interesting cases will require significantly
improved accuracy in the simulations. Given that the present simulations
already utilize significant computer resources, rather than increase
the resolution a more realistic near term solution would be to
incorporate higher-order finite difference techniques or spectral 
methods (\cite{utb,utb2} already use 
fourth order differencing
in space and time, and~\cite{nasa,nasa2} use fourth order in space and second order
in time). Modulo the accuracy issues, numerical relativity finally seems
to have entered the era where it will begin uncover what will hopefully be
the very rich and interesting landscape of black hole
interactions.

\noindent{\bf{\em Acknowledgments:}}
FP gratefully acknowledges research support from CIAR.
The simulations described here were performed on 
the University of British Columbia's {\bf vnp4}
cluster (supported by CFI and BCKDF), {\bf WestGrid} machines
(supported by CFI, ASRI and BCKDF), and Dell {\bf Lonestar} cluster
at the University of Texas in Austin.

\end{document}